\documentclass[12pt]{iopart}
\usepackage{graphicx} 
\begin{document}

\def\la{\langle}
\def\ra{\rangle}
\newcommand{\beq}{\begin{equation}}
\newcommand{\eeq}{\end{equation}}
\newcommand{\beqa}{\begin{eqnarray}}
\newcommand{\eeqa}{\end{eqnarray}}

\title[Gaussian wave packet impinging on a square barrier]{Explicit solution for a Gaussian
wave packet impinging on a square barrier}

\author{A. L. P\'erez Prieto*, S. Brouard*, and J. G. Muga\dag}
\address{*Departamento de F\'\i sica Fundamental II,
Universidad de La Laguna, Spain}
\address{\dag Departamento de Qu\'\i mica-F\'\i sica, Universidad del
Pa\'\i s Vasco, Apdo. 644, 48080 Bilbao, Spain}

\begin{abstract}
The collision of a quantum Gaussian wave packet with a square barrier
is solved explicitly in terms of known functions. The obtained formula
is suitable for performing fast calculations or asymptotic analysis. 
It also provides physical insight since the description of different
regimes and collision phenomena typically requires only some of the
terms.   
\end{abstract}
\maketitle
%

%

\section{Introduction}

A paradigmatic textbook example \cite{Schiff} of a quantum scattering process 
is the collision of a particle of mass $m$, initially represented by a
minimum-uncertainty-product Gaussian wave packet, with a
``square barrier'' potential, 
\beq
V(x)=\cases{V_0 \quad &
${\rm if}\;\;-d/2\le\,x\,\le\,d/2$
\cr
0 \quad &${\rm otherwise.}$\cr}
\eeq
In spite of its simplicity, this model allows to observe and study
a number of interesting quantum phenomena, such as tunneling
\cite{Huang89, Leavens90, BSM94, KBK96, XA97, AAGT98, AG00,PR00},
resonances \cite{BC70}, incidence-reflection and
incidence-transmission interferences \cite{PBM01},
the Hartman effect \cite{Hartman62, BSM94, DM96,PMB97,Ruschhaupt98,Ruschhaupt00},
and time delays or reflection due to a well when
the height of the potencial, $V_0$, is negative
\cite{GSS67}. It is also used
as a standard model to check the validity of approximate
numerical methods \cite{TS87, Muga91}, or to exemplify
and test different theories for temporal quantities such
as arrival, dwell, or decay times, the asymptotic
behaviour at long times \cite{LJPU00,BEM01,tqm}, and
quantum transition state theories \cite{MDSS96}. 

Except for approximate analytical treatments \cite{Hartman62},
the solution of this time-dependent scattering model has been 
always obtained by numerical methods, typically using
``Fast Fourier Transforms'', tridiagonal systems, or by linear
combination of the solutions of the stationary Schr\"odinger
equation. Goldberg, Schey, and Schwartz were the first to solve
the model by means of a difference equation \cite{GSS67}.
While for many purposes these methods may be sufficient, for 
other applications the numerical approach may be cumbersome,
time consuming or even useless. Therefore, an analytical
solution in terms of known functions is of much practical interest.     
In general an analytical solution is not only useful as a fast
computational tool; it is also valuable because of the physical
insight than can be gained from it, very often by means of
approximations that extract the dominant contributions in different
limits, parameter ranges and regimes.  
   
In this paper we present an explicit expression in the momentum
representation of the time dependence of a Gaussian packet incident
on a square barrier in terms of known functions. There are a few other
time dependent solvable scattering models: a Lorentzian state impinging
on a delta function potential \cite{EK88,MBS92,MDS95} or on a separable exponential
potential \cite{MDS95,MP98}; and a cutoff plane wave impinging on a square
barrier \cite{BM96,GCV01}, or on a step barrier \cite{DCM02}.
While all have been useful and illustrative
of several scattering properties, the present model is the only one that
combines simultaneously a local potential, resonances, and a physical
Hilbert-space initial state with finite moments.

\section{Explicit solution}
 
The initial wave function is assumed to be a minimum-uncertainty-product
Gaussian wave packet located in the left half-line and with negligible
overlap with the potential. In momentum representation it is given by   
\beq\label{eig}
\la p'|\psi(t=0)\ra=\left(\frac{2\delta_x}{\pi\hbar^2}\right)^{1/4}
\,\exp\left[-\delta_x\left(p'-p_c\right)^2/\hbar^2-
ip'x_c/\hbar\right],
\eeq
where $x_c$ and $p_c$ are the initial mean position and momentum
respectively, and $\delta_x$ is the variance of the state in coordinate
representation. Its time evolution can then be written in terms of
eigenstates of $H$, $\left|\phi_{p'}\right>$ \cite{HWZ77,BEM02}, as    
\beq\label{sdt}
\psi(p,t)=\int_{-\infty}^{+\infty}\,\la p|\phi_{p'}\ra e^{-iE't/\hbar}
\la p'|\psi(t=0)\ra \,dp',
\eeq
with $E'\equiv p'^2/2m$. These states are given explicitly in coordinate
representation by  
\beq\label{xp'}
\la x|\phi_{p'}\ra=\frac{1}{h^{1/2}}\cases{
Ie^{ip'x/\hbar}+Re^{-ip'x/\hbar}& $\qquad x<-d/2$\cr
Ce^{ip''x/\hbar}+De^{-ip''x/\hbar}&$\qquad -d/2\le x\le d/2$\cr
Te^{ip'x/\hbar}&$\qquad x>d/2$,\cr} 
\eeq
where $p''\equiv{\sqrt{p^{'2}-2mV_0}}$. The value of the coefficient 
$I$ will be taken as
$1$. 
Note that these (delta-normalized) eigenstates are associated with an incident
plane wave of momentum $p'$ for $p'>0$ and with an outgoing plane wave with
momentum $p'$ for $p'<0$. The coefficients $R$, $C$, $D$ and $T$ are determined
by continuity of the wave function and its first derivative,   
\beqa
&&R\left(p'\right)=\frac{i\,\left(\frac{p''}{p'}-
\frac{p'}{p''}\right)\sin\left(p''d/\hbar\right)
e^{-ip'd/\hbar}}{2\Omega\left(p'\right)}\label{coeb}\nonumber\\
&&C\left(p'\right)=\frac{\left(1+\frac{p'}{p''}\right)
e^{-i(p'+p'')d/2\hbar}}{2\Omega\left(p'\right)}\nonumber\\
&&D\left(p'\right)=\frac{\left(1-\frac{p'}{p''}\right)
e^{-i(p'-p'')d/2\hbar}}{2\Omega\left(p'\right)}\nonumber\\
&&T\left(p'\right)=\frac{e^{-ip'd/\hbar}}{\Omega\left(p'\right)},
\label{coef}
\eeqa
where 
\beq
\Omega\left(p'\right)\equiv\cos(p''d/\hbar)-
\frac{i}{2}\left(\frac{p''}{p'}+\frac{p'}
{p''}\right)\sin(p''d/\hbar).
\label{omega}
\eeq
Fourier transformation of $\la x|\phi_{p'}\ra$ in  (\ref{xp'})
%
%
gives five terms proportional to the coefficients 
$I,R,T,C,D$ respectively (generically denoted by $A$ hereafter). 
Substituting these terms into  (\ref{sdt}) 
with the initial state in (\ref{eig}), we obtain  
%
%
\beq\label{iabf}
\psi (p,t)=i\tau\hbar h^{-1/2}\int_{-\infty}^{\infty}
\left[g_I\left(p'\right)+
g_R\left(p'\right)+g_C\left(p'\right)+g_D\left(p'\right)+
g_T\left(p'\right)\right]
e^{\phi\left(p'\right)}dp'\,,
\eeq
where $\tau\equiv\left(2\pi\hbar\right)^{-1/2}\left(\frac{2\delta_x}
{\pi\hbar^2}\right)^{1/4}$. The term in the exponential $\phi(p')$ is
given by  
\beq\label{phi}
\phi\left(p'\right)=\frac{-i{p'}^2t}{2m\hbar}-\frac{\delta_x\left(p'-
p_c\right)^2}{\hbar^2}-\frac{ip'x_c}{\hbar}
-\frac{ip'd}{2\hbar};
\eeq
and 
%
\beqa
g_I(p')&\equiv&\frac{e^{ipd/2\hbar}}{p-p'+i\hbar 0^+}
\nonumber\\
g_R(p')&\equiv&\frac{R(p')e^{ip'd/\hbar}\,e^{ipd/2\hbar}}{p+p'+i\hbar 0^+}
\nonumber\\
g_C(p')&\equiv&\frac{-2iC(p')e^{ip'd/2\hbar}}{p-p''}\sin[(p-p'')d/2\hbar]
\nonumber\\
g_D(p')&\equiv&\frac{-2iD(p')e^{ip'd/2\hbar}}{p+p''}\sin[(p+p'')d/2\hbar]
\nonumber\\
g_T(p')&\equiv&\frac{-T(p') e^{ip'd/\hbar}\,e^{-ipd/2\hbar}}{p-p'-i\hbar 0^+}.
\eeqa
Note the three explicit (``structural'' \cite{WS95}) 
poles at $p'_I\equiv p+i\hbar 0^+$,
$p'_T\equiv p-i\hbar 0^+$, and $p'_R\equiv -p-i\hbar 0^+$, in addition to the poles of the
functions $C$, $D$, $R$ and $T$, $p'_j$ ($j=1,...,\infty$), which are zeros of $\Omega(p')$.
All the poles lie in the lower half complex plane except for the incidence pole, $p'_I$.

The integral in Eq. (\ref{iabf}) can be solved by completing the square
in Eq.(\ref{phi}) and introducing the variable $u$ as  
\beq
u\equiv\frac{p'-z}{f},
\eeq
with
\beq
f\equiv\left(\frac{\delta_x}{\hbar^2}+i\frac{t}{2m\hbar}\right)^{-1/2}.
\eeq
$u$ is zero at the saddle point $z$, defined by 
\beq
z\equiv\frac{m\left[4mp_c\delta_x^2-\left(x_c+d/2\right)\hbar^2
t\right]-i2m\hbar\left[m\delta_x\left(x_c+d/2\right)+p_c
\delta_xt\right]}{4m^2\delta_x^2+t^2\hbar^2},
\eeq
and becomes real along the steepest descent path, a straight line with
slope $-t\hbar/(2m\delta_x)$. We now deform the contour and integrate
along the steepest descent path, namely along the real $u$ axis,   
%
\beqa
\label{psiabfu}
\psi(p,t)&=&i f \tau\hbar h^{-1/2}\,
e^{-\left(\delta_x p_c^2/\hbar^2\right)
+\eta^2}
\nonumber\\
&\times&\int_{\Gamma_u}\left[g_I\left(u\right)+g_R\left(u\right)+
g_C\left(u\right)
+g_D\left(u\right)+
g_T\left(u\right)\right]e^{-u^2}\,du,
\eeqa
where 
\beq
\eta\equiv\left(\frac{2p_c\delta_x}{\hbar^2}-i\frac{\left(x_c+d/2
\right)}{\hbar}\right)\left[4\left(\frac{\delta_x}{\hbar^2}+
i\frac{t}{2m\hbar}\right)\right]^{-1/2}, 
\eeq
$g_A(u)\equiv g_A(p'=fu+z)$, and $\Gamma_u$ goes from $-\infty$ to
$+\infty$ including a circle
around the poles that have been crossed by the contour deformation.
Since the $g_I$, $g_R$, $g_T$ and $g_{CD}\equiv g_C+g_D$ are
meromorphic functions with simple poles, it is useful to extract
explicitly the singularities and leave the remainder as an entire
function, $h$,  
%
\beqa
&&g_I(u)=\frac{{\cal R}_I}{u-u_I}
\nonumber\\
&&g_R(u)=\frac{{\cal R}_R}{u-u_R}+\sum_{j=1}^{\infty}\,
\frac{{\cal R}_{Rj}}{u-u_{j}}+h_R(u)
\nonumber\\
&&g_{CD}(u)=\sum_{j=1}^{\infty}\,\frac{{\cal R}_{Cj}+{\cal R}_{Dj}}{u-u_{j}}+
h_{CD}(u)
\nonumber\\
&&g_T(u)=\frac{{\cal R}_T}{u-u_T}+\sum_{j=1}^{\infty}\,
\frac{{\cal R}_{Tj}}{u-u_{j}}+h_T(u). 
\label{gbf}
\eeqa
Again the poles in the $u$ complex plane may be separated into ``structural''
\cite{WS95},     
\beqa
u_I&=&f^{-1}(p+i0^+-z)\nonumber\\
u_R&=&f^{-1}(-p-i0^+-z)\nonumber\\
u_T&=&f^{-1}(p-i0^+-z)\label{ues},
\eeqa
and ``resonance'' poles (see the Appendix), 
\beq
u_j=f^{-1}(p'_j-z)\qquad\qquad j=1,...,\infty.
\eeq
The residues ${\cal R}$ are given by  
%
\beqa
&&{\cal R}_I=-f^{-1}\,e^{ipd/2\hbar}
\nonumber\\
&&{\cal R}_R=f^{-1}\,R(p'_R) e^{ip'_R d/\hbar}\,e^{ipd/2\hbar}
\nonumber\\
&&{\cal R}_{Rj}=\frac{R(p'_j)e^{ip'_j d/\hbar}
\,e^{ipd/2\hbar} F_j}{\left(p+p'_j+i\hbar 0^+\right)f}
\nonumber\\
&&{\cal R}_{Cj}=\frac{-2iC(p'_j)
e^{ip'_jd/2\hbar}\sin[(p-p''_j)d/2\hbar]F_j}
{\left(p-p''_j\right)f}
\nonumber\\
&&{\cal R}_{Dj}=\frac{-2iD(p'_j)e^{ip'_jd/2\hbar}
\sin[(p+p''_j)d/2\hbar] F_j}
{\left(p+p''_j\right)f}
\nonumber\\
&&{\cal R}_T=f^{-1}\,T(p'_T) e^{ip'_Td/\hbar}\,e^{-ipd/2\hbar}
\nonumber\\
&&{\cal R}_{Tj}=-\frac {\,T(p'_j)e^{ip'_jd/\hbar}e^{-ipd/2\hbar}F_j}{\left(p-p'_j-
i\hbar 0^+\right)f},
\eeqa
where 
\beq
F_j\equiv\frac{\Omega(p')}{d\Omega(p')/dp'}\Big|_{p'=p'_j}.
\eeq
Taking into account the integral expression of the $w$-function, 
$w(z)=e^{-z^2}\,erfc(-iz)$ \cite{Abram}, 
\beq\label{wz}
w(z)=\frac{1}{i\pi}\int_{\Gamma_-} \,\frac{e^{-u^2}}{u-z}\, du, 
\eeq
where $\Gamma_-$ goes from $-\infty$ to $\infty$ passing below the pole,  
and the relation $w(-z)=2e^{-z^2}-w(z)$, the wave function may finally
be written as 
%
%
\beqa
\psi{(p,t)}&=&if\tau\hbar h^{-1/2}\,e^{-\left(\delta_x p_c^2/\hbar^2\right)
+\eta^2}\Big\{i\pi {\cal R}_I w(u_I)-i\pi {\cal R}_R
w(-u_R)
\nonumber\\
&-&i\pi {\cal R}_T w(-u_T)
-i\pi \sum_{j=1}^{\infty}\, [{\cal R}_{Rj}+{\cal R}_{Cj}
+{\cal R}_{Dj}+{\cal R}_{Tj}]
 w(-u_j)
\nonumber\\
&+&\int_{-\infty}^{\infty}\,[h_R(u)+h_C(u)+h_D(u)+h_T(u)]e^{-u^2}du
\Big\},
\label{mr}
\eeqa
which is the main result of this paper. The remaining Gaussian integrals
are quite generally only a minor correction and may be evaluated with a
few terms of the series
\beq
h(u)=\sum_{n=0}^{\infty}\,
h^{(n)}(0)\frac{u^n}{n!}, 
\eeq
which gives
\beq
\label{hs}
\int_{-\infty}^{\infty}h(u)e^{-u^2}du=
\sqrt{\pi}\left[h(u=0)+\sum_{n=1}^{\infty}\,\frac{1\times 3\times...
\times(2n-1)}{2^n(2n)!}\,h^{(2n)}(u=0)\right].
\eeq
Note the form of the solution in (\ref{mr}). There are structural terms 
associated with incidence, transmission and reflection ($I$, $T$, and $R$
terms for short), resonance terms, and the $h$-corrections of (\ref{hs}).

\section{Examples}

\begin{figure}
\begin{center}
{\includegraphics[angle=-90,width=3in]{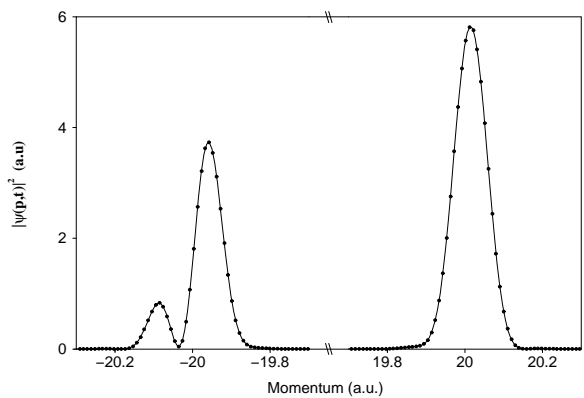}}
\end{center}
\caption{$|\psi(p,t)|^2$ as a function of momentum at $t=5$ a.u.,
when the collision
has been completed. The solid line
corresponds to the exact solution, whereas the dots correspond
to the reflection and transmission
(structural) terms and three resonance pole terms from (\ref{mr}). Atomic units
are used in all the calculations, and $m=1$. $p_c=20$, $d=2.5$, $V_0=188$,
$\delta_x=100$ and $x_c$ is located $50$ atomic units to the left of the center
of the barrier potential.
}
\label{fig1}
\end{figure}
\begin{figure}
\begin{center}
{\includegraphics[angle=-90,width=3in]{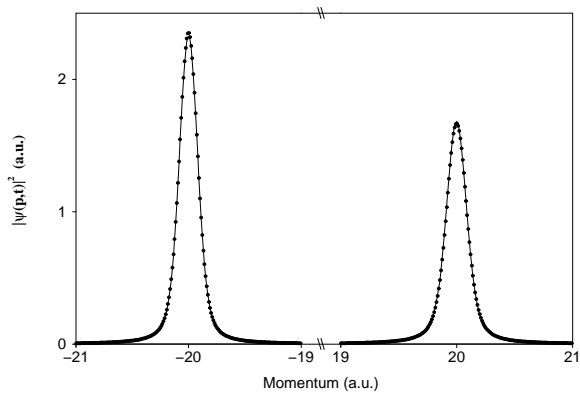}}
\end{center}
\caption{$|\psi(p,t)|^2$ as a function of momentum,
at $t=2.5$ a.u., during the
collision process. The solid line
corresponds to the exact solution whereas the dots correspond
to incidence and reflection structural terms of (\ref{mr}) 
only, and no resonance poles taken into account.
$p_c=20$, $d=3$, $V_0=400$, $\delta_x=100$, 
and the initial wave packet is centered $50$ atomic units
to the left of the barrier center.
}
\label{fig2}
\end{figure}
\begin{figure}
\begin{center}
{\includegraphics[angle=-90,width=3in]{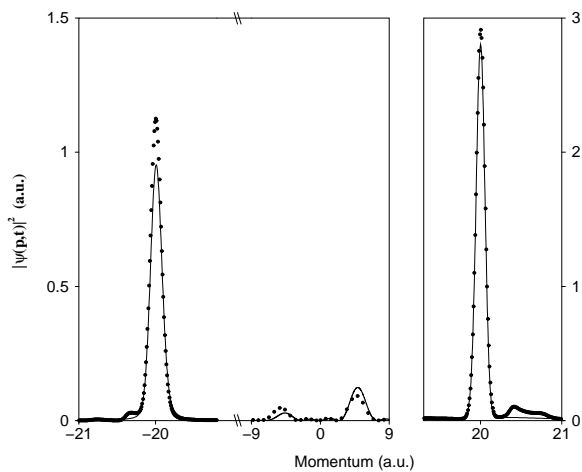}}
\end{center}
\caption{$|\psi(p,t)|^2$ as a function of momentum,
at $t=2.7$ a.u., during the
collision process. The solid line
corresponds to the exact solution, whereas the dots correspond
to taking from (\ref{mr}) incidence, reflection, transmission,
and barrier ($C$ and $D$) terms, plus three resonance pole terms.
When one correction term (\ref{hs}) is added,
the approximate result is indistinguishable
from the exact one.
Same barrier and wave packet parameters as in Fig. 1
}
\label{fig3}
\end{figure}

In many cases just a few terms in
(\ref{mr}) provide an accurate approximation to the exact result.  
If only the asymptotic behaviour, after the collision has been completed, is
of interest, the structural terms $R$ and $T$, associated with reflection 
and transmission, plus a few resonance pole contributions (again for
reflection and transmission) are quite sufficient. The number of resonance
poles that have to be added depends on the energy of the collision. In
figure 1, for example, we show a case where only three resonance poles
have been used in the $R$ and $T$ terms, and no correction term from Eq.
(\ref{hs}) has been included.

There are also cases where only two terms are required {\em during} the 
collision: the collision with a very opaque barrier ($R$ and $I$ terms),
see Fig. 2,  and collisions for very broad an energetic wave packets
($I$ and $T$ terms). This later case has been used recently to explain
a transient, classically forbidden enhancement of the high momentum
components \cite{BM98a,BM98b} as an
incidence-transmission interference effect \cite{PBM01}. 

In Figure 3, the case of a very general (typical) situation is depicted.
Incidence, reflection, transmission,
and barrier ($C$ and $D$) terms from (\ref{mr}), plus three resonance
pole terms provide a fairly good
approximation for the wave function at an intermediate time during the
collision, although the approximate solution clearly overestimates the exact value
at the reflection and transmission peaks, and for values of momentum 
between $20.3$ and $21$ a.u.. The inclusion of the first correction term in (\ref{hs}) leads to 
a curve which is indistinguishable in the scale of the figure from
the exact result.

Let us also note that (\ref{mr}) is suitable for asymptotic analysis
(e.g. short and large times) by means of the asymptotic series of the
$w$-function, see e.g. \cite{BM96, MP98}. The $w$-functions may be
considered the elementary propagators of the Schr\"odinger transient modes
\cite{Nus} and play a prominent role in the explicit solution given in
(\ref{mr}).  

\ack{
AP acknowledges support by CajaCanarias; 
SB acknowledges support by the Ministerio de Ciencia y Tecnolog\'\i a of Spain
(Project BFM2001-3349);  
JGM acknowledges support 
by the Ministerio de Ciencia y Tecnolog\'\i a (BFM2000-0816-C03-03), 
UPV-EHU (00039.310-13507/2001), and the Basque Government (PI-1999-28).}

\appendix
\section{Calculation of the resonance poles}
The resonance poles are obtained by solving the transcendental 
equation $\Omega(p')=0$. A systematic way to do it is 
based on the differential equation
\cite{simtodim}
\beq\label{ecdif}
\frac{dp'}{dV_0}=\frac{idV_0\sqrt{m}-p'}{V_0\left(
\frac{idp'}{\sqrt{m}}-2\right)}.
\eeq
As boundary
condition for the integration of (\ref{ecdif}) we take $p'=0$
at 
\beq\label{init}
V_0=-\frac{n^2\pi^2}{2d^2};\qquad\qquad n=0,\pm 1,\pm 2,...
\eeq
The integration of (\ref{ecdif}) has to be carried out until the
value of $V_0$ corresponds to the physical barrier we want to examine. 
The integration however cannot be performed along the real
axis because of the factor $V_0$ in the denominator of (\ref{ecdif}).
To avoid the singularity at $V_0=0$, an imaginary term is added to the 
potential, analytically continuing the
differential equation.  A parabolic path has been used across the complex 
$V_0$ plane from the initial values of $V_0$ in (\ref{init}) to the final
real value of $V_0$.     


\end{document}